\documentclass[12pt,a4paper]{article}
\usepackage{mathptmx}
\usepackage{amsmath}
\usepackage{amssymb}
\usepackage{amstext}
\usepackage{color}
\usepackage{graphicx}
\usepackage{url}
\usepackage{float}
\usepackage{sidecap}

\usepackage{subfig}
\usepackage{listings}
\newcommand{\likwid}{LIKWID}

\newcommand{\bq}{\begin{equation}}
\newcommand{\eq}{\end{equation}}

\newcommand{\flop}{\mbox{flop}}
\newcommand{\flops}{\mbox{flops}}

\newcommand{\GBS}{\mbox{GB/s}}

\newcommand{\GFS}{\mbox{GFlop/s}}

\newcommand{\GHZ}{\mbox{GHz}}

\newcommand{\bytes}{\mbox{bytes}}

\newcommand{\MB}{\mbox{MB}}

\newcommand{\eos}{~.}

\begin{document}
\begin{center}\LARGE
Parallel sparse matrix-vector multiplication
as a test case for hybrid MPI+OpenMP programming
\\[0.5cm]\large
Gerald Schubert$^1$, Georg Hager$^1$, Holger Fehske$^2$, Gerhard Wellein$^{1,3}$\\[1mm]\normalsize\itshape
${}^1$Erlangen Regional Computing Center, 91058 Erlangen, Germany\\
${}^2$Institute for Physics, University of Greifswald, 17487 Greifswald, Germany\\
${}^3$Department of Computer Science, University of Erlangen-Nuremberg, \\91058 Erlangen, Germany
\end{center}
\begin{abstract}
  We evaluate optimized parallel sparse matrix-vector operations for
  two representative application areas on widespread multicore-based
  cluster configurations.  First the single-socket baseline
  performance is analyzed and modeled with respect to basic
  architectural properties of standard multicore chips. Going beyond
  the single node, parallel sparse matrix-vector operations often
  suffer from an unfavorable communication to computation
  ratio. Starting from the observation that nonblocking MPI is not
  able to hide communication cost using standard MPI implementations,
  we demonstrate that explicit overlap of communication and
  computation can be achieved by using a dedicated communication
  thread, which may run on a virtual core. 
  We compare our approach to pure MPI and the
  widely used ``vector-like'' hybrid programming strategy.  
\end{abstract}

\section{Introduction}

Many problems in science and engineering involve the solution of large
eigenvalue problems or extremely sparse systems of linear equations
arising from, e.g., the discretization of partial differential
equations. Sparse matrix-vector multiplication (spMVM) is the dominant
operation in many of those solvers and may easily consume most of the
total run time.
A highly efficient scalable spMVM implementation is thus fundamental,
and complements advancements and new development in the high-level
algorithms.

For more than a decade there
has been an intense debate about whether the hierarchical structure
of current HPC systems needs to be considered in parallel programming, 
or if pure MPI is
sufficient. Hybrid approaches based on MPI+OpenMP have
been implemented in codes and kernels for various applications areas
and compared with traditional MPI implementations. Most results are
hardware-specific, and sometimes contradictory.
In this paper we analyze hybrid MPI+OpenMP variants
of a general parallel spMVM operation. Beyond the naive
approach of using OpenMP for parallelization of kernel
loops (``vector mode'') we also employ a hybrid
``task mode''  to overcome or mitigate a weakness of
standard MPI implementations: the lack of
truly asynchronous communication in nonblocking
MPI calls. 
We test our
implementation against pure MPI approaches for two application
scenarios on an InfiniBand cluster as well as a Cray XE6 system.

\subsection{Related work}

In recent years the performance of various spMVM algorithms has been
evaluated by several groups~\cite{h0441_GKAKK08,h0441_WOVSYD09,h0441_BG09p}.
Covering different matrix storage formats and implementations on
various types of hardware, they reviewed a more or less large number
of publicly available matrices and reported on the obtained
performance.
Scalable parallel spMVM implementations have also been
proposed \cite{symspmvm10,spmvmgr01}, mostly
based on an MPI-only strategy. Hybrid parallel spMVM approaches have
already been devised before the emergence of multicore
processors~\cite{rw03,vecpar02}. Recently a ``vector mode'' approach could not
compete with a scalable MPI implementation for a specific
problem on a Cray system~\cite{symspmvm10}. There is no up-to-date
literature that systematically investigates novel features like
multicore, ccNUMA node structure, and simultaneous multithreading
(SMT) for hybrid parallel spMVM.

\subsection{Sparse matrix-vector multiplication and node-level performance model}
\label{sect:perfmod}

A matrix is called ``sparse'' if the number of its nonzero entries
grows only linearly with the matrix dimension. Since keeping such a
matrix in computer memory with all zero entries included is out of the
question, an efficient format to store the nonzeros only is required.
The most widely used variant is ``Compressed Row Storage''
(CRS)~\cite{barrett:1994}.
It does not exploit specific features that
may emerge from the underlying physical problem like, e.g., 
block structures, symmetries, etc., but is broadly
recognized as the most efficient format 
for general sparse matrices on cache-based microprocessors. All nonzeros are
stored in one contiguous array \verb.val., row by row, and the
starting offsets of all rows are contained in a separate array
\verb.row_ptr.. Array \verb.col_idx. contains the original column
index of each matrix entry.
A matrix-vector multiplication with a RHS vector \verb.B(:).
and a result vector \verb.C(:). can then be written as follows:
\begin{lstlisting}
  do i = 1,%$N_\mathrm{r}$%
    do j = row_ptr(i), row_ptr(i+1) - 1
      C(i) = C(i) + val(j) * %B(col\verb._.idx(j))%
    enddo
  enddo
\end{lstlisting}
Here $N_\mathrm r$ is the number of matrix rows. While arrays \verb.C(:).
and \verb.val(:). are traversed contiguously, access to \verb.B(:). is
indexed and may potentially cause very low spatial and temporal
locality in this data stream.


The performance of spMVM operations on a single compute
node is often limited by main memory 
bandwidth. Code balance~\cite{hpc4se} is thus a good metric to establish a simple
performance model. We assume the average length of the inner ($j$) loop to
be $N_\mathrm{nzr}=N_\mathrm{nz}/N_\mathrm r$, 
where $N_\mathrm{nz}$ is the total number of
nonzero matrix entries. Then the contiguous data
accesses in the CRS code generate $(8+4+16/N_\mathrm{nzr})$ \bytes{} of 
memory traffic for a single inner loop iteration, where the first two contributions come
from the matrix \verb.val(:). (8\,\bytes) and the index array
\verb.col_idx(:). (4\,\bytes), while the last term reflects the
update of \verb.C(i). (write allocate + evict). 
The indirect access pattern to \verb.B(:). is determined by the
sparsity structure of the matrix and can not be modeled in general. 
However, \verb.B(:). needs to be loaded at least once from main
memory, which adds another $8/N_\mathrm{nzr}$\,\bytes{} per inner
iteration. Limited cache size and nondiagonal access typically require
loading at least parts of \verb.B(:). multiple times in a single MVM. 
This is quantified by a machine- and problem-specific parameter $\kappa$:
For each additional time that \verb.B(:). is loaded from main memory,
$\kappa=8/N_\mathrm{nzr}$ additional bytes are needed.
Together with the
computational intensity of 2\,\flops\ per $j$ iteration the code balance is
\bq\label{eq:pmodel}
B_\mathrm{CRS}=\left(\frac{12 + 24/N_\mathrm{nzr} + \kappa}{2}\right)\frac{\bytes}{\flop}
=\left(6+\frac{12}{N_\mathrm{nzr}}+\frac{\kappa}{2}\right) \frac{\bytes}{\flop}\eos
\eq
On the node level $B_\mathrm{CRS}$ can be used to determine an upper
performance limit by measuring the node memory bandwidth (e.g., using 
the STREAM benchmark) and assuming
$\kappa=0$. Moreover, from the sparse MVM floating point performance
and the memory bandwidth drawn by the CRS code, $\kappa$ can be
determined experimentally (see Sect.~\ref{sec:nodeperf}). Since
the matrices used here have $N_\mathrm{nzr} \approx 7\ldots 15$, 
each additional access to
\verb.B(:). incurs a nonnegligible contribution to the data
transfer.

Note that this simple model neglects performance-limiting aspects
beyond bandwidth limitations, like load imbalance, communication
and/or synchronization overhead, and the adverse effects of
nonlocal memory access across ccNUMA locality domains (LDs).

\subsection{Experimental setting}

\subsubsection{Test matrices}

Since the
sparsity pattern may have substantial impact on the single
node performance and parallel scalability, we have chosen two
application areas known to generate extremely sparse matrices. 

As a first test case we use a matrix from exact diagonalization
of strongly correlated electron-phonon systems in solid state
physics. Here generic
microscopic models are used to treat both charge (electrons) and
lattice (phonons) degrees of freedom in second quantization. Choosing
a finite-dimensional basis set, which is the direct product of basis
sets for both subsystems (electrons $\otimes$ phonons), the generic
model can be represented by a sparse Hamiltonian matrix. 
Iterative algorithms such as Lanczos or Jacobi-Davidson are used to
compute low-lying eigenstates of the Hamilton matrices, and more recent
methods based on polynomial expansion 
allow for computation of spectral properties~\cite{WWAF06} or time
evolution of quantum states~\cite{WF06}. In all those algorithms,
sparse MVM is the most time-consuming step.

\begin{figure}[tbp]
\includegraphics*[width=\textwidth]{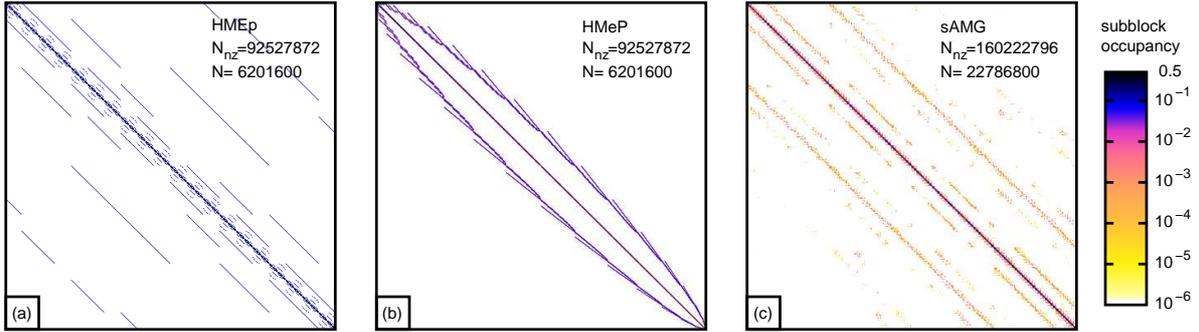}
\caption{Sparsity patterns of the Hamiltonian matrix
 described in the text with different numbering of the basis
 elements ((a) and (b)), and the sAMG matrix (c). Square subblocks
 have been aggregated and color-coded according to occupancy to improve visibility.
}\label{fig:mat}
\end{figure}

In this paper we consider the Holstein-Hubbard model
(cf.~\cite{fwhwb04} and references therein) and choose six electrons
(subspace dimension 400) on a six-site lattice coupled to 15 phonons
(subspace dimension $1.55\times10^4$). The resulting matrix of
dimension $6.2\times 10^6$ is very sparse ($N_\mathrm{nzr}\approx 15$)
and can have two different sparsity patterns, depending on whether the
phononic or the electronic basis elements are numbered contiguously
(see Figs.~\ref{fig:mat}\,(a) and (b), respectively). 
We also applied the well-known ``Reverse Cuthill-McKee (RCM)''
algorithm~\cite{RCM69} to the Hamilton matrix  in order to improve spatial locality  in the access to the right hand side vector,
and to optimize interprocess communication patterns towards near-neighbor
exchange.
Since the RCM-optimized
structure showed no performance advantage over the HMeP variant (Fig.~\ref{fig:mat}\,(b))
neither on the node nor on the highly parallel level,
we will not consider RCM any further here.

The second matrix is generated by the adaptive multigrid code sAMG
(see~\cite{AMG,sAMG} and references therein) for the irregular discretization
of a Poisson problem on a car geometry. Its matrix dimension is
$2.2\times10^7$ with an average of $N_\mathrm{nzr}\approx 7$ entries
per row (see Fig.~\ref{fig:mat}\,(c)).


For real-valued, symmetric matrices as considered here it is
sufficient to store the upper triangular matrix elements and perform,
e.g., a parallel symmetric CRS sparse MVM~\cite{symspmvm10}. The data
transfer volume is then reduced by almost a factor of two, allowing for
a corresponding performance improvement. We do
not use this optimization here for two major reasons. First, the
discussion of the hybrid parallel vs. MPI-only implementation should
not be restricted to the special case of explicitly symmetric
matrices. Second, to our knowledge an efficient shared memory
implementation of a symmetric CRS sparse MVM base routine has not yet been
presented.

\subsubsection{Test machines}

\paragraph{Intel Nehalem EP / Westmere EP}

\begin{figure}[tbp]
\subfloat[Intel dual Westmere node with two NUMA locality domains]{\includegraphics[height=0.32\linewidth]{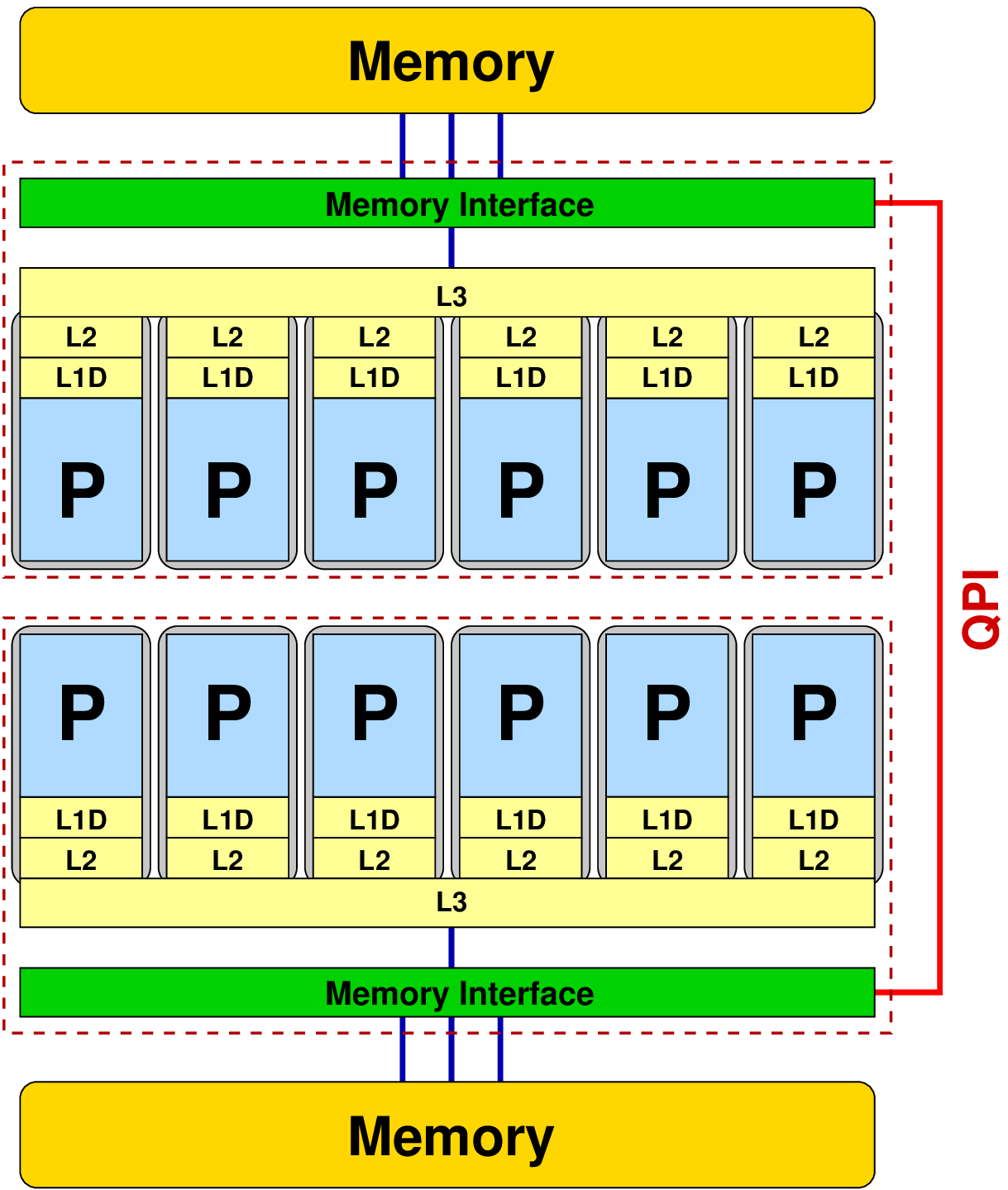}}\hfill
\subfloat[Cray XE6/AMD dual Magny Cours node with four NUMA locality domains]{\includegraphics[height=0.32\linewidth]{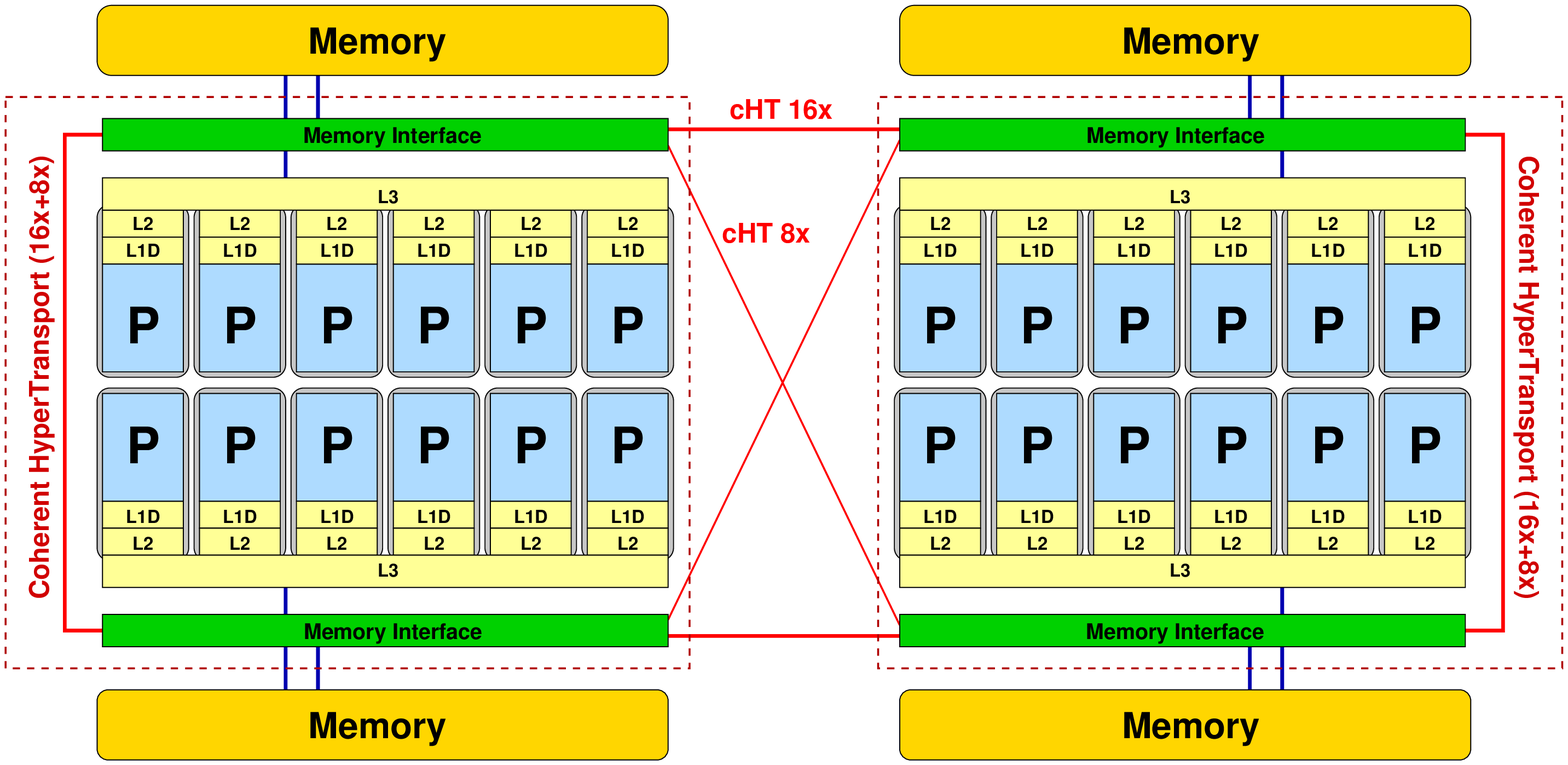}}
\caption{\label{fig:hardware}Node topology of the benchmark systems. Dashed boxes indicate sockets.}
\end{figure}

The two Intel platforms
represent a ``tick'' step within Intel's ``tick-tock'' product
strategy. Both processors only differ in a few microarchitectural
details; the most important difference is that Westmere, due
to the 32\,nm production process, accommodates
six cores per socket instead of four while keeping the same
L3 cache size per core (2\,\MB) as Nehalem. 
The processor chips (Xeon X5550 and X5650) used for the benchmarks
run at 2.66\,\GHZ{} base frequency with
``Turbo Mode'' and Simultaneous Multithreading (SMT) enabled.
A single socket forms its own ccNUMA LD via  three
DDR3-1333 memory channels (see Fig.~\ref{fig:hardware}\,(a)), allowing
for a peak bandwidth of 32\,\GBS. We use standard dual-socket nodes
that are connected via fully nonblocking QDR InfiniBand (IB)
networks. The Intel compiler in version 11.1 and the Intel MPI library
in version 4.0.1 were used throughout. Thread-core affinity was
controlled with the \likwid~\cite{likwid} toolkit.

\paragraph{Cray XE6 / AMD Magny Cours}

The Cray XE6 system is based on dual-socket nodes with AMD Magny Cours 
12-core processors (2.1\,\GHZ{} Opteron 6172) and the latest Cray ``Gemini''
interconnect. The internode bandwidth of the 2D torus network is 
beyond the capability of QDR InfiniBand.  The single node architecture depicted in
Fig.~\ref{fig:hardware}(b) reveals a unique feature of the AMD Magny
Cours chip series: The 12-core package comprises two 6-core chips
with separate L3 caches and memory controllers, tightly bound by
``1.5'' HyperTransport (HT) 16x links. Each 6-core unit forms its own NUMA
LD via two DDR3-1333 channels, i.e.,
a two-socket node comprises four NUMA locality domains.
In total the AMD design uses eight memory
channels, 
allowing for a theoretical main memory bandwidth advantage of $8/6$
over a Westmere node.  The Cray compiler in version 7.2.8 was used for
the Cray/AMD measurements. 

\section{Node-level performance analysis}\label{sec:nodeperf}

The basis for each parallel program must be an efficient single
core/node implementation. For general sparse matrix structures the CRS
format presented above is very suitable for modern
cache-based multicore processors~\cite{SHF09}. Even advanced 
machine-specific optimizations such as nontemporal prefetch
instructions for Opteron
processors provide only minor benefits~\cite{symspmvm10} and are thus
not considered here. A simple OpenMP parallelization of the
outermost loop, together with an appropriate NUMA-aware data
placement strategy has proven to provide best
node-level performance. We choose the HMeP
matrix as a reference problem. The results presented
hold qualitatively for the other matrix structures as well.
Differences will be discussed where required.

\begin{figure}[tbp]
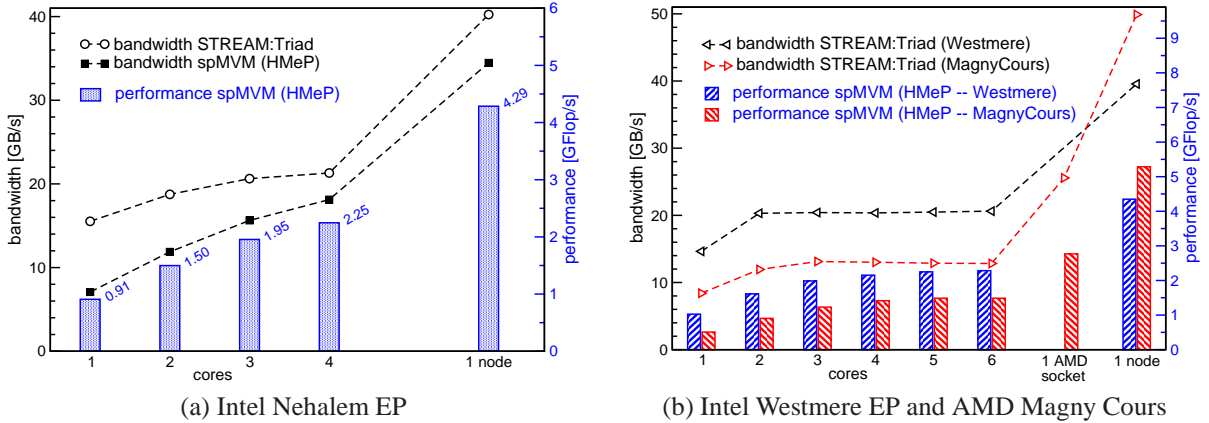

\subfloat[Intel Nehalem EP]{\includegraphics*[width=0.48\textwidth]{Neha_Node}}\hfill
\subfloat[Intel Westmere EP and AMD Magny Cours]{\includegraphics*[width=0.48\textwidth]{MaCoWest_Node}}\hfill
\caption{Node-level performance for the test systems. Effective STREAM 
triads bandwidth$^1$, and performance for 
sparse MVM using the HMeP matrix (bars) is shown. In (a) we also report the 
measured memory bandwidth for the sparse MVM operation.}\label{fig:nodeperf}
\end{figure}

Intrasocket and intranode spMVM scalability
should always be discussed together with effective STREAM triads
numbers, which form a practical upper bandwidth 
limit.\footnote{Nontemporal stores have been suppressed in 
the STREAM measurements  and the bandwidth numbers reported have been scaled
appropriately ($ \times 4/3$) to account for the write-allocate transfer.} 
On the Nehalem EP platform, the memory bandwidth drawn by the 
spMVM as measured with \likwid~\cite{likwid} is also shown 
in Fig.~\ref{fig:nodeperf}\,(a).
While the STREAM bandwidth soon saturates within a socket, the spMVM
bandwidth and the corresponding \GFS\ numbers still benefit from
the use of all cores. This is a typical behavior for codes with 
(partially) irregular data access patterns. However, the fact that
more than 85\% of the STREAM bandwidth can be reached with spMVM
indicates that our CRS implementation makes good use of the
resources. The maximum spMVM performance can be estimated by dividing
the memory bandwidth by the code balance (\ref{eq:pmodel}), using $N_\mathrm{nzr}=15$
and $\kappa=0$. For a single socket the spMVM draws 18.1\,\GBS{}
(STREAM triads: 21.2\,\GBS{}), allowing for a maximum performance of
2.66\,\GFS{} (3.12\,\GFS{}). 
Combining the measured performance (2.25\,\GFS{}) and
bandwidth of the spMVM operation with $B_\mathrm{CRS}(\kappa)$ we find
$\kappa=2.5$, i.e., 2.5 additional \bytes\ of memory traffic on \verb.B(:).
per inner loop iteration  (37.3\,\bytes\ per row) are 
required due to limited cache capacity. Thus the complete vector \verb.B(:). is
loaded six times from main memory to cache, but each element is used
$N_\mathrm{nzr}=15$ times. This ratio
gets worse if the matrix bandwidth increases. For the HMEp matrix we found
$\kappa=3.79$, which translates to a 50\% increase in the additional
data transfers for \verb.B(:).. The code balance implies a performance
drop of about 10\%, which is consistent with our measurements.

In Fig.~\ref{fig:nodeperf}\,(b) we summarize the performance
characteristics for Intel Westmere and AMD Magny Cours, which
both comprise six cores per locality domain. While the
AMD system is weaker on a single LD, its node-level performance
is about 25\% higher than on Westmere due to its four LDs
per node. Within the domains spMVM saturates at four cores on
both architectures, leaving ample room to use the remaining cores for
different tasks, like communication (see Sect.~\ref{sec:taskmode}).
In the following we will report results for the Westmere and Magny Cours 
platforms only.

\section{Distributed-memory parallelization}\label{sec:dmpar}

Strong scaling of MPI-parallel sparse MVM is inevitably limited
by communication overhead. Hence, it is vital to find ways to
hide communication costs as far as possible. A widely used
approach is to employ nonblocking point-to-point MPI calls
for overlapping communication with useful work. 
However, it has been known for a long time that most MPI
implementations support progress, i.e., actual data transfer, only 
when MPI library code is executed by the user process, although the
hardware even on standard InfiniBand-based clusters does
not hinder truly asynchronous point-to-point communication. Using the
simple benchmark from \cite{hpc4se} we have
verified that this situation has not changed with
current MPI versions (Intel 4.0.1, OpenMPI 1.5). In the following
sections we will contrast the ``naive'' overlap applying nonblocking MPI
with an approach that uses a dedicated OpenMP thread for 
explicitly asynchronous transfers. We adopt the nomenclature 
from~\cite{rw03,hpc4se} and distinguish between ``vector mode''
and ``task mode.''

%
%

\subsection{Vector-like parallelization: Vector mode}\label{sec:vectormode}

\begin{figure}
\subfloat[Vector mode, no overlap]{\includegraphics[width=0.32\linewidth]{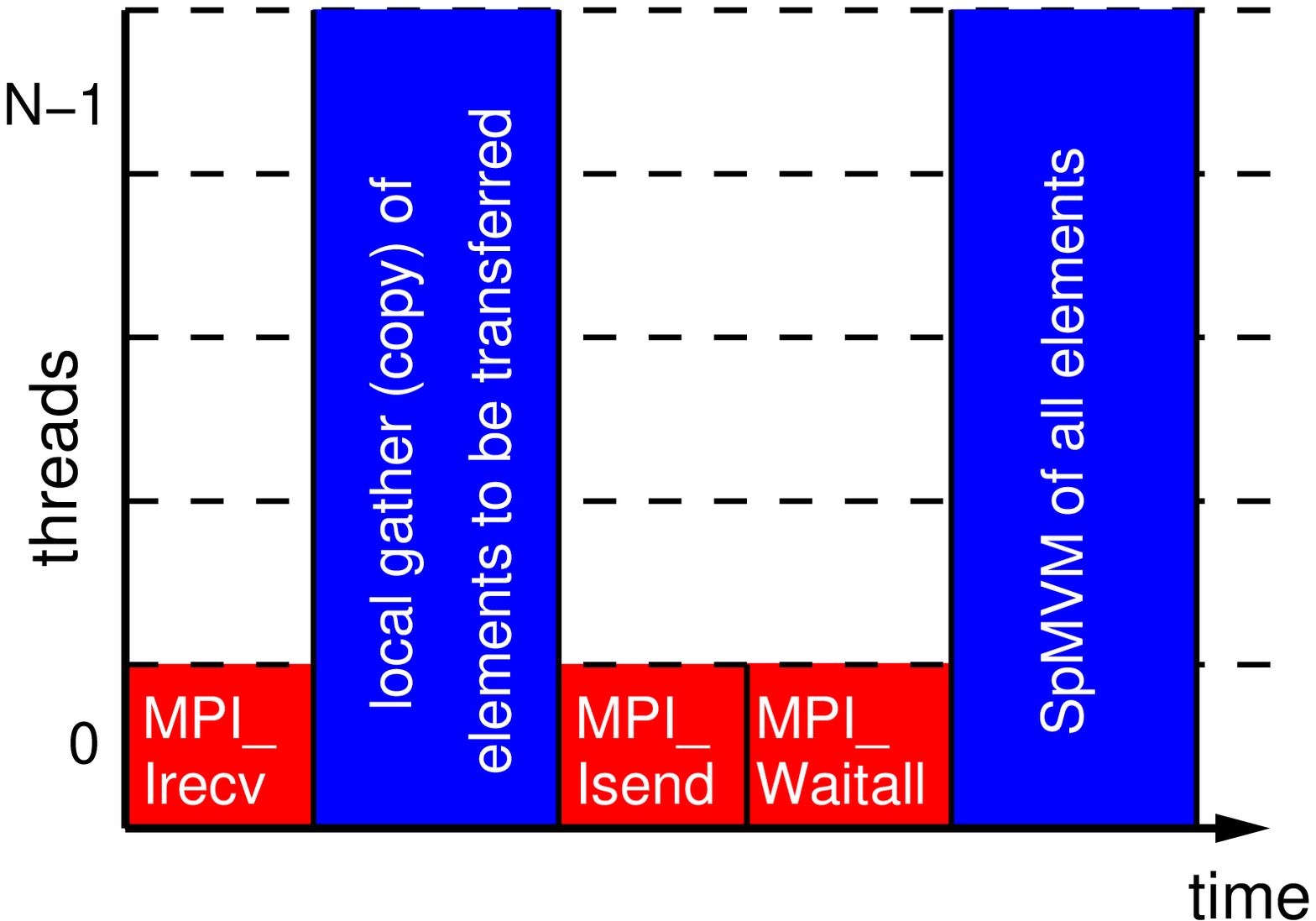}}\hfill
\subfloat[Vector mode, naive overlap]{\includegraphics[width=0.32\linewidth]{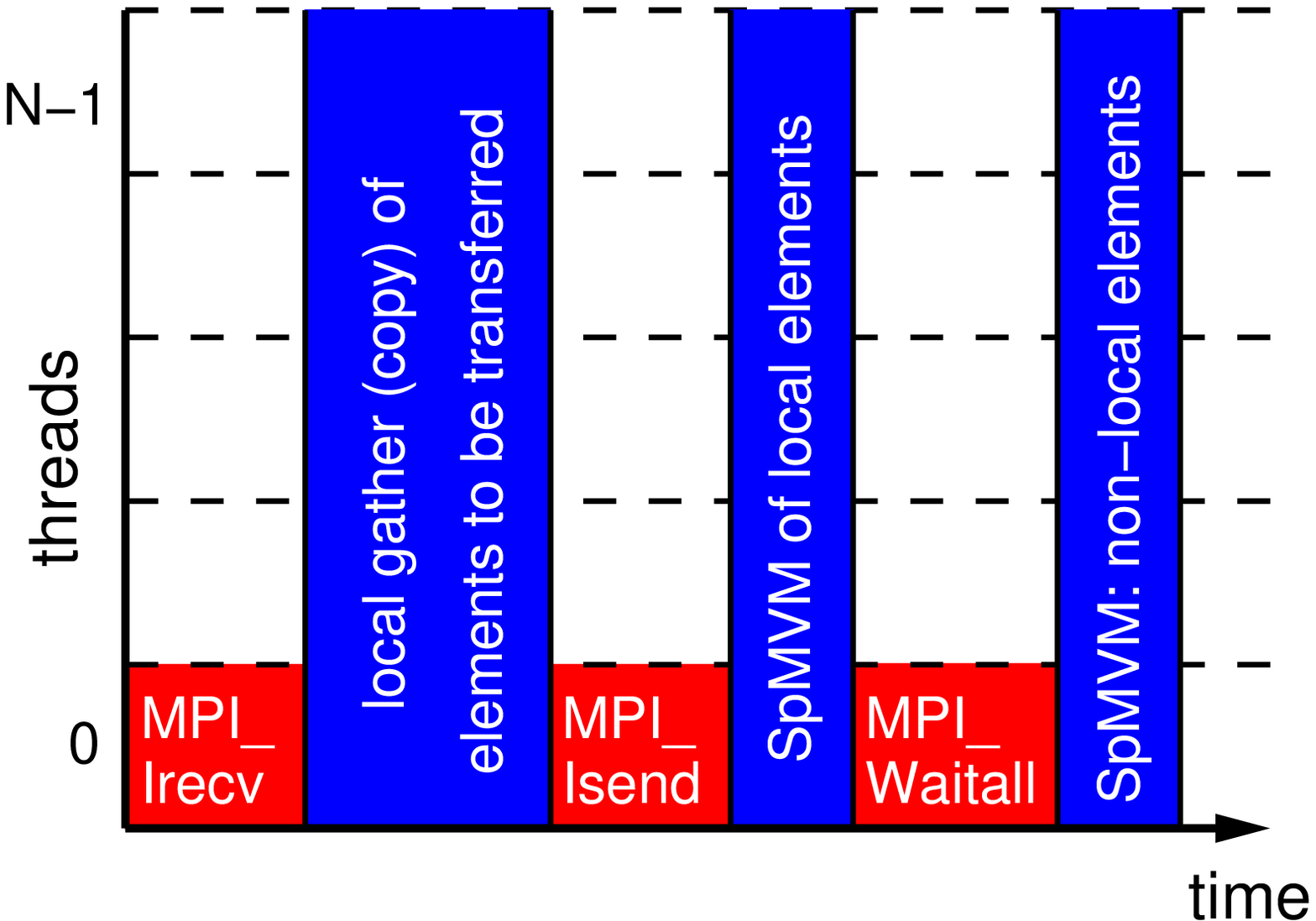}}\hfill
\subfloat[Task mode, explicit overlap]{\includegraphics[width=0.32\linewidth]{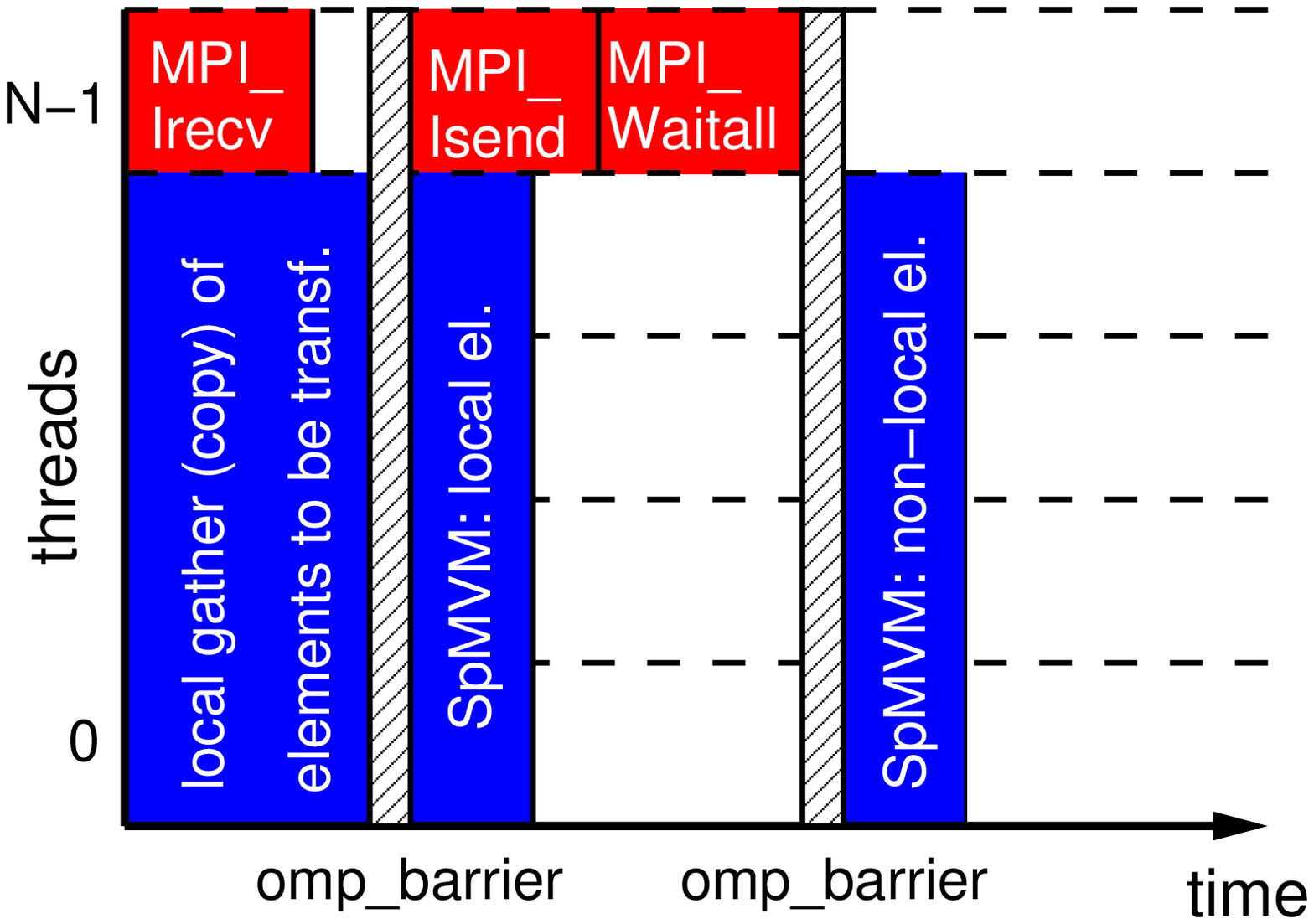}}
\caption{\label{fig:kernel_schemes}Schematic timeline view of the implemented hybrid kernel versions. From 
left to right: no communication/calculation overlap, naive overlap using nonblocking MPI, 
and explicit overlap by a dedicated communication thread}
\end{figure}
MPI parallelization of spMVM is generally done by distributing the
nonzeros (or, alternatively, the matrix rows), the right hand side
vector \verb.B(:)., and the result vector \verb.C(:). evenly across
MPI processes. Due to off-diagonal nonzeros, every process requires
some parts of the RHS vector from other processes to complete its own
chunk of the result, and must send parts of its own RHS chunk to
others.\footnote{Note that it is generally difficult to 
establish good load balancing for computation and communication
at the same time. We use a balanced distribution of nonzeros
across the MPI processes here.} The resulting communication pattern depends only on the
sparsity structure, so the necessary bookkeeping needs to be done only
once. After the communication step is over, the local spMVM can be
performed, either by a single thread or, if threading is available, by
multiple threads inside the MPI process.  Gathering the data to be
sent into a contiguous send buffer may be done after the receive has
been initiated, potentially hiding the cost of copying (see
Fig.~\ref{fig:kernel_schemes}\,(a)). We call this naive approach
``hybrid vector mode,'' since it strongly resembles the
programming model for vector-parallel computers~\cite{rw03}: The time-consuming
(although probably parallel) computation step does not overlap with
communication overhead.  This is actually how ``MPI+OpenMP hybrid
programming'' is still defined in most publications. The question whether
and why using multiple threads per MPI process may improve performance
compared to a pure MPI version on the same hardware is not easy to
answer, and there is no general rule.

As an alternative one may consider hybrid vector mode with nonblocking
MPI (see Fig.~\ref{fig:kernel_schemes}\,(b)) to potentially overlap
communication with the part of spMVM that can be completed using local
RHS elements only.  After the nonlocal elements have been received,
the remaining spMVM operations can be performed. A disadvantage of
splitting the spMVM in two parts is that the local result vector must
be written twice, incurring additional memory traffic. The performance
model (\ref{eq:pmodel}) can be modified to account for an additional data transfer of
$16/N_\mathrm{nzr}$ bytes per inner loop iteration, leading to
a modified code balance of 
\bq\label{eq:mpmodel}
B^\mathrm{split}_\mathrm{CRS}
=\left(6+\frac{20}{N_\mathrm{nzr}}+\frac{\kappa}{2}\right) \frac{\bytes}{\flop}\eos
\eq
For $N_\mathrm{nzr}\approx 7\ldots 15$ and assuming $\kappa=0$,
one may expect a node-level performance penalty between
15\% and 8\%, and even less if $\kappa>0$.

For simplicity we will also use the term ``vector mode'' for pure
MPI versions with single-threaded computation.

\subsection{Explicit overlap of communication and computation: Task mode}\label{sec:taskmode}

A safe way to ensure overlap of communication with computation is to
use a separate communication thread and leave the computational loops
to the remaining threads. We call this ``hybrid task mode,'' because
it establishes a functional decomposition of tasks (communication vs.
computation) across the resources (see
Fig.~\ref{fig:kernel_schemes}\,(c)): One thread executes MPI calls
only, while all others are used to copy data into send buffers,
perform the spMVM with the local RHS elements, and finally (after all
communication has finished) do the remaining spMVM parts.  Since spMVM
saturates at about 3--5 threads per locality domain (as shown in
Fig.~\ref{fig:nodeperf}\,(b)), at least one core per LD is available
for communication without adversely affecting node-level performance.
On architectures with SMT, like the Intel Westmere, there is also the
option of using one compute thread per physical core and bind the
communication thread to a logical core.

Apart from the additional memory traffic due to writing the result
vector twice (see Sect.~\ref{sec:vectormode}), another drawback of
hybrid task mode is that the standard OpenMP loop worksharing
directive cannot be used, since there is no concept of ``subteams'' in
the current OpenMP standard. Work distribution is thus implemented explicitly,
using one contiguous chunk of nonzeros per compute thread.

\section{Performance results and discussion}\label{sec:performance}

\begin{figure}\centering
\includegraphics[width=0.95\linewidth,clip]{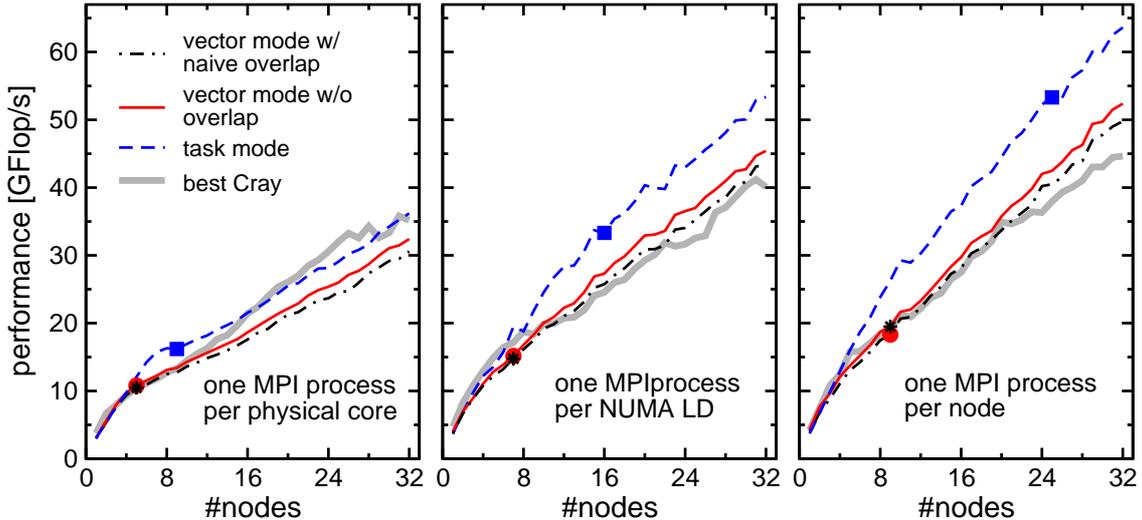}
\caption{\label{fig:nodescaling_rrze3vv}Strong scaling performance data for
  spMVM with the HMeP matrix on the Intel Westmere cluster for
  different pure MPI and hybrid variants. The 50\% parallel efficiency
  point with respect to the best single-node version is indicated
  on each data set. The best variant on the Cray XE6 system is shown
  for reference.}
\end{figure}

\begin{figure}\centering
\includegraphics[width=0.95\linewidth,clip]{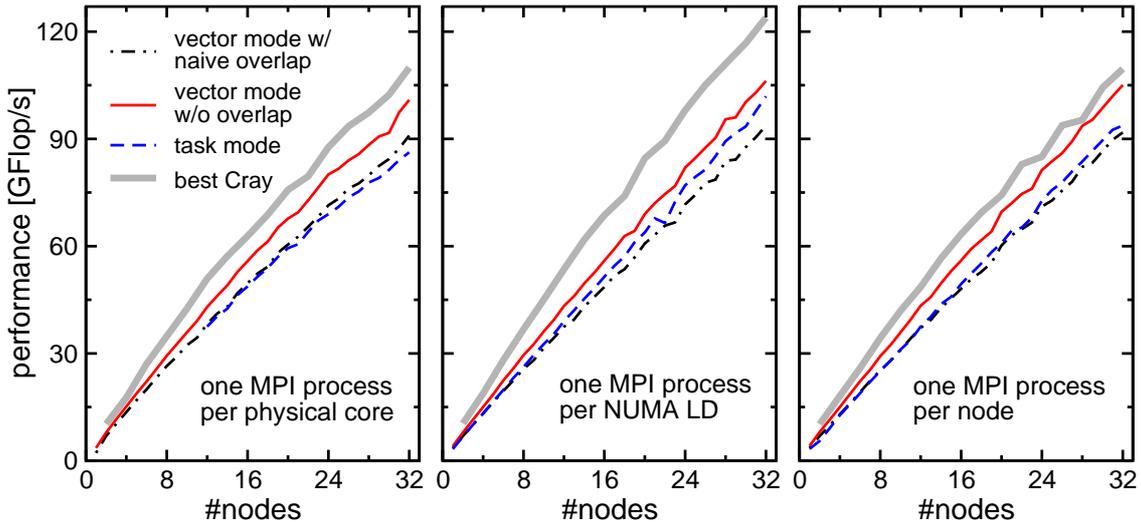}
\caption{\label{fig:nodescaling_scai2}Strong scaling performance data for
  spMVM with the sAMG matrix (same variants as in 
  Fig.~\ref{fig:nodescaling_rrze3vv}). Parallel efficiency is above 50\%
  for all versions up to 32 nodes. The Cray system performed best 
  in vector mode without overlap for all cases.}
\end{figure}

Figures~\ref{fig:nodescaling_rrze3vv} and \ref{fig:nodescaling_scai2} show strong
scaling results for the two chosen matrices (HMeP and sAMG) and different
parallelization schemes on the Westmere cluster. For HMeP 
(Fig.~\ref{fig:nodescaling_rrze3vv}) we have indicated the
50\% parallel efficiency point (with respect to the best single-node
performance as reported in Fig.~\ref{fig:nodeperf}\,(b)) on each data set;
in practice one would not go beyond this number of nodes because of 
bad resource utilization. 

\paragraph{HMeP} (see Fig.~\ref{fig:nodescaling_rrze3vv})
At one MPI process per physical core (left panel), 
vector mode with naive overlap is always slower than the variant without overlap because the
additional data transfer on the result vector cannot be compensated by
overlapping communication with computation.  Task mode was implemented
here with one communication thread per MPI process, running on the
second virtual core. In this case, point-to-point transfers explicitly
overlap with the local spMVM, leading to a noticeable performance
boost.  One may conclude that MPI libraries with support for progress
threads could follow the same strategy and bind those threads to
unused logical cores, allowing overlap even with single-threaded
user code.

With one MPI process per NUMA locality domain (middle panel)
the advantage of task mode is even more pronounced. Since
the memory bus of an LD is already saturated with four threads,
it does not make a difference whether six worker threads are
used with one communication thread on a virtual core, or whether
a physical core is devoted to communication. The same is
true with only one MPI process (12 threads) per node (right
panel). For the matrix and the system under investigation
it is clear that task mode allows strong scaling to much 
higher levels of parallelism with acceptable parallel 
efficiency than any variant of vector mode. 

Contrary to expectations based on the single-node performance numbers
(Fig.~\ref{fig:nodeperf}\,(b)), the Cray XE6 can generally not match
the performance of the Westmere cluster at larger node counts, with
the exception of pure MPI where both are roughly on par. We have 
observed a strong influence of job topology and machine load on the
communication performance over the 2D torus network. Since sparse MVM
requires significant non-nearest-neighbor communication with 
growing process counts, the nonblocking fat tree network on the Westmere
cluster seems to be better suited for this kind of problem.

Interestingly, the hybrid vector mode variants with one MPI process
per LD or per node already provide better scalability than pure MPI;
we attribute this to the smaller number of messages in the hybrid 
case (message aggregation) and a generally improved load balancing.
There is also a universal drop in scalability beyond about six
nodes, which is largely independent of the particular hybrid
mode. This can be ascribed to a strong decrease in overall
internode communication volume when the number of nodes is small.
The effect is somewhat less pronounced for pure MPI, since the overhead
of intranode message passing cannot be neglected.

\paragraph{sAMG} (see Fig.~\ref{fig:nodescaling_scai2})
The sAMG matrix has much weaker communication requirements than HMeP,
and the impact of load imbalance is very small. Hence, all variants
and hybrid modes (pure MPI, one process per LD, and one process per
node) show similar scaling behavior and there is no advantage of task
mode over naive, pure MPI without overlap. This situation supports
the general rule that it makes no sense to consider MPI+OpenMP hybrid 
programming if the pure MPI code already scales well and behaves 
in accordance with a single-node performance model.

On the Cray XE6, vector mode without overlap performs best across
all hybrid modes, with a significant advantage of running one
MPI process with six threads per LD. This aspect is still to be 
investigated.

\section{Summary and outlook}

We have investigated the performance properties of different pure MPI
and MPI+OpenMP hybrid variants of sparse matrix-vector multiplication
on two current multicore-based parallel systems, using two matrices
with significantly different sparsity patterns.  The single-node
performance model and analysis on Intel Westmere and AMD Magny Cours
processors showed that memory-bound sparse MVM saturates the memory
bus of a NUMA locality domain already at about four threads, leaving
free resources for implementing explicit computation/communication
overlap. Since most current standard MPI implementations do not
support truly asynchronous point-to-point communication, explicit
overlap enabled substantial performance gains in strong scaling
scenarios for communication-bound problems, especially when running
one process per NUMA domain or per node. As the communication thread
can run on a virtual core, MPI implementations could use the same
strategy for internal ``progress threads'' and so enable asynchronous
communication without changes in MPI-only user code.

Future work will cover a more complete investigation of load balancing
effects, and a careful analysis of the performance properties
of the Cray XE6 system. We will also employ development versions
of MPI libraries that support asynchronous progress and compare 
with our hybrid task mode approach.

\section*{Acknowledgments}
We thank J.~Treibig, R.~Keller and T.~Sch{\"o}nemeyer for valuable
discussions, A.~Basermann for providing the RCM transformation,
and K.~St{\"u}ben and H.\,J.~Plum for providing and supporting the AMG test
case. We acknowledge financial support from KONWIHR II (project
HQS@HPC II) and thank CSCS Manno for providing access to their Cray
X6E system\@.

\end{document}